\documentstyle[12pt]{article}

\newcommand{\EQ}{\begin{equation}}
\newcommand{\EN}{\end{equation}}
\newcommand{\bea}{\begin{eqnarray}}
\newcommand{\ena}{\end{eqnarray}}
\newcommand{\vs}[1]{\vspace{#1 mm}}

\renewcommand{\b}{\beta}
\renewcommand{\c}{\gamma}

\newcommand{\shalf}{\frac{1}{2}}
\newcommand{\pa}{\partial}
\newcommand{\dz}{\frac{dz}{2\pi i}}
\newcommand{\uda}{\nearrow \kern-1em \searrow}

\begin{document}

\topmargin 0pt
\oddsidemargin 5mm

\renewcommand{\Im}{{\rm Im}\,}
\newcommand{\NP}[1]{Nucl.\ Phys.\ {\bf #1}}
\newcommand{\PL}[1]{Phys.\ Lett.\ {\bf #1}}
\newcommand{\CMP}[1]{Comm.\ Math.\ Phys.\ {\bf #1}}
\newcommand{\PR}[1]{Phys.\ Rev.\ {\bf #1}}
\newcommand{\PRL}[1]{Phys.\ Rev.\ Lett.\ {\bf #1}}
\newcommand{\PTP}[1]{Prog.\ Theor.\ Phys.\ {\bf #1}}
\newcommand{\PTPS}[1]{Prog.\ Theor.\ Phys.\ Suppl.\ {\bf #1}}
\newcommand{\MPL}[1]{Mod.\ Phys.\ Lett.\ {\bf #1}}
\newcommand{\IJMP}[1]{Int.\ Jour.\ Mod.\ Phys.\ {\bf #1}}

\begin{titlepage}
\setcounter{page}{0}
\begin{flushright}
OS-GE 37-93\\
hep-th/9310180
\end{flushright}

\vs{15}
\begin{center}
{\Large BOSONIZATION OF A TOPOLOGICAL COSET MODEL \\
AND NON-CRITICAL STRING THEORY}
\vs{15}

{\large Nobuyoshi Ohta$^{1,2}$ and Hisao Suzuki$^{2}$}\\
\vs{8}
$^1${\em NORDITA, Blegdamsvej 17, DK-2100 Copenhagen \O, Denmark}\\
\vs{5}
$^2${\em Institute of Physics, College of General Education,
Osaka University \\ Toyonaka, Osaka 560, Japan} \\
\end{center}
\vs{10}

\centerline{{\bf{Abstract}}}

We analyze the relation between a topological coset model based on
super $SL(2,R)/U(1)$ coset and non-critical string theory by using
free field realization. We show that the twisted $N=2$ algebra of
the coset model can be naturally transformed into that of non-critical
string. The screening operators of the coset models can be identified
either with those of the minimal matters or with the cosmological
constant operator. We also find that another screening operator,
which is intrinsic in our approach, becomes the  BRST nontrivial
state of ghost number $0$ (generator of the ground ring for $c=1$
gravity).

\end{titlepage}
\newpage
\renewcommand{\thefootnote}{\arabic{footnote}}
\setcounter{footnote}{0}

The relation between non-critical strings and topological field
theories is the subject of current interest. It has long been suggested
that the latter theories describe the unbroken phase of
gravity~\cite{WIT1}, but their precise relation has not been clear.

It has been known that the twisting of $N=2$ superconformal field
theory gives rise to topological theory~\cite{WIT1,EY}. This suggests
that any non-critical string theories may have hidden $N=2$
superconformal symmetry. Indeed, several authors have observed that
the BRST current and the antighost field $b(z)$ generate an algebra
that is quite similar but apparently not identical to the $N=2$
superconformal algebra~\cite{DVF}. It turns out that the BRST current
can be modified by total derivative terms so that the antighost and
the physical BRST current exactly generate a topologically twisted
$N=2$ superconformal algebra~\cite{GS,BLNW}. This does not identify,
however, the structure of the models with $N=2$ symmetry.

Recently, rather nontrivial correspondence between super
$SL(2,R)_k/U(1)$ coset model~\cite{KS} and $c=1$ string has been analyzed
through twisted $N=2$ structure. Mukhi and Vafa~\cite{MV} have
revealed an amazing correspondence between these two models for $k=3$.
In this letter, we discuss the relation of these models and the
generalization of the correspondence to the minimal models coupled to
gravity by means of the free field realization. We find that there is
another interesting correspondence for $k=1$.

Super $SL(2,R)_k / U(1)$ model is described by the bosonic coset
model of $SL(2,R)_k \times U(1)/U(1)$~\cite{DPZ}. For a representation of
$SL(2,R)_k$, we use the following free field realization~\cite{WAK}:
\bea
J^+ &=& \frac{1}{\sqrt{k-2}} \left[ \b \c^2 - \sqrt{2(k-2)} \c \pa
  \phi_3 + k \pa \c \right], \nonumber\\
J^3 &=& \b \c - \sqrt{\frac{k-2}{2}} \pa \phi_3, \nonumber\\
J^- &=& \sqrt{k-2} \b,
\ena
where the basic fields $(\b, \c)$ with the dimension $(1,0)$
and $\phi_3$ satisfy
\EQ
\b(z) \c(w) \sim { 1 \over z-w },
\qquad \phi_3(z) \phi_3(w) \sim - \ln(z-w).
\EN
We use fermions ${\bar\psi}$ and $\psi$ to represent a $U(1)$ element.
Then  $N=2$ algebra can be realized by means of these fields as follows
\cite{DPZ}:
\bea
T &=& \frac{1}{k-2}\left[\shalf\{ J^+,J^-\}-(J^3)^2\right]
 -\shalf{\bar\psi}\pa\psi-\shalf\psi\pa{\bar\psi}
 +\frac{1}{k-2}(J^3-\psi{\bar\psi})^2 \nonumber\\
&=&\b \pa \c - \shalf (\pa \phi_3)^2 + \frac{1}{\sqrt{2(k-2)}}
 \pa^2 \phi_3 \nonumber\\
& &- \shalf {\bar\psi}\pa \psi - \shalf \psi \pa {\bar\psi}
 + \frac{1}{k-2}\left( \b\c - \sqrt{\frac{k-2}{2}} \pa \phi_3
 - \psi{\bar\psi}\right)^2, \nonumber\\
G^+ &=& \frac{1}{\sqrt{k-2}} \psi J^+
 = \frac{1}{k-2}\psi \left[ \b\c^2 - \sqrt{2(k-2)}\c \pa \phi_3
 + k \pa \c \right], \nonumber\\
G^- &=& \frac{1}{\sqrt{k-2}} {\bar\psi} J^-
 = {\bar\psi}\b, \nonumber\\
J &=& \frac{1}{k-2}( k\psi{\bar\psi} - 2J^3).
\ena
The screening operators of $SL(2,R)$ are given by
\EQ
V_- = \b e^{\sqrt{\frac{2}{k-2}} \phi_3},
  \qquad V_+ = \b^{k-2} e^{\sqrt{2(k-2)} \phi_3}.
\EN

We would like to identify $G^-$ with the antighost $b$ of the non-critical
string. For this
purpose, it turns out to be convenient to bosonize $(\b, \c)$
and $({\bar\psi}, \psi)$ fields as
\bea
\b &=& e^{\phi_1 + i\phi_2}, \qquad
\c = -i \pa \phi_2 e^{- \phi_1 -i\phi_2}, \nonumber\\
{\bar\psi} &=& e^{- i\sigma}, \qquad
\psi = e^{i\sigma},
\ena
where fields $\phi_1, \phi_2$ and $\sigma$ are normalized to have
the ordinary correlations, e. g. $\phi_1(z)\phi_1(w) \sim - \ln (z-w)$.
Notice that this is a little different from the usual bosonization
rule~\cite{FMS}. We should also note that we have an additional screening
operator $S(z)$ in this particular bosonization:
\EQ
S(z) = e^{i \phi_2(z)}.
\EN
In terms of the free bosons $\phi_1, \phi_2, \phi_3$ and $\sigma$,
the generators of $N=2$ elements are written as
\bea
T &=& - \shalf ( \pa \phi_1)^2 - \shalf( \pa \phi_2)^2 +
 \shalf \pa^2 \phi_1 - \frac{i}{2} \pa^2 \phi_2 \nonumber\\
& & - \shalf( \pa \phi_3)^2 + \frac{1}{\sqrt{2(k-2)}}
 \pa^2 \phi_3 - \shalf (\pa \sigma)^2 \nonumber\\
& & + \frac{1}{k-2}\left( \pa \phi_1 - \sqrt{\frac{k-2}{2}} \pa\phi_3
 - i\pa \sigma\right)^2, \nonumber\\
G^+ &=& \frac{1}{k-2}\left[i(k-2) \pa \phi_1 \pa \phi_2 -
 (k-1)(\pa \phi_2)^2 \right. \nonumber\\
& & \left.+ i \sqrt{2(k-2)} \pa\phi_2\pa\phi_3 -i(k-1)\pa^2\phi_2\right]
 e^{i\sigma - \phi_1 -i\phi_2}, \nonumber\\
G^- &=& e^{-i\sigma + \phi_1 + i\phi_2} \nonumber\\
J &=& \frac{1}{k-2}\left[ ki\pa \sigma -
 2\left( \pa \phi_1 - \sqrt{ k-2 \over 2} \pa \phi_3\right)\right].
\ena
In these expressions, we have three characteristic combinations of
bosons. The first is the one lying in the direction of $N=2$
$U(1)$ charge, which we call $u$:
\EQ
iu = i\sqrt{ k \over k-2} \sigma -{ 2 \over \sqrt{k(k-2)}}
\phi_1 + \sqrt{ 2 \over k} \phi_3.
\EN
The second represents the direction eliminated by the coset construction:
\EQ
\sqrt{k-2 \over 2} \rho \equiv - i\sigma + \phi_1 -
\sqrt{ k-2 \over 2} \phi_3.
\EN
The third one is in the direction of the screening charge of $SL(2,R)$:
\EQ
\sqrt{2 \over k-2} \phi \equiv \phi_1 + i \phi_2 + \sqrt{2 \over k-2}
 \phi_3.
\EN
We finally define the remaining combination out of four bosons in
such a way that it is orthogonal to the other three combinations:\footnote{
The factors of $i$ are introduced in the above field redefinitions in such
a way that all fields have the ordinary correlations. We define the field
$\phi_2$ anti-hermitian not to mix the hermiticity of the fields.}
\EQ
i\sqrt{ 2 \over k-2} \tau= \sqrt{ k-2 \over k} \phi_1 +
i \sqrt{ k \over k-2} \phi_2 + \sqrt{ 2 \over k}\phi_3 .
\EN
The generators of $N=2$ algebra can be expressed by these bosons as
\bea
T &=& - { 1 \over 2} (\pa u)^2 - { 1\over2} (\pa \phi)^2 -
{1\over2} (\pa \tau)^2 + { k-1 \over \sqrt{2(k-2)}} \pa^2 \phi -
\sqrt{k \over 2} i \pa^2 \tau, \nonumber\\
G^- &=& e^{-\sqrt{ k-2 \over k}iu + \sqrt{2 \over k}i\tau}, \nonumber\\
J&=& \sqrt{k \over k-2}i \pa u.
\ena

We are now going to twist this system to topological conformal
models~\cite{EY}
by defining the energy-momentum tensor as $ T \to T + {1 \over 2}\pa J$.
By defining BRST charge $Q_B = \oint \dz G^+(z)$, we can find
that energy-momentum tensor is BRST trivial: $T(z) = \{ Q_B , G^-(z)\}$.
The twisted energy-momentum tensor of this model is given by
\EQ
T = - \shalf (\pa u)^2 - \shalf (\pa \phi)^2 - \shalf (\pa \tau)^2
+ { k-1 \over \sqrt{2(k-2)}} \pa^2 \phi - i\sqrt{k \over 2}
 \pa^2 \tau + \frac{i}{2} \sqrt{ k \over k-2} \pa^2 u.
\EN
To make a connection with $c\leq 1$ gravity, let us make our final
field redefinition. Since $G^-$ is given by
a simple expression, it is natural to fermionize this as
\EQ
b \equiv G^- = e^{-\sqrt{ k-2 \over k}iu + \sqrt{2 \over k}i\tau}, \qquad
c \equiv e^{\sqrt{ k-2 \over k}iu - \sqrt{2 \over k}i\tau}.
\EN
 Defining the other orthogonal combination of fields $u$ and $\tau$ as
\EQ
X \equiv \sqrt{2 \over k} u + \sqrt{k-2 \over k}\tau,
\EN
we finally get the elements of the topological algebra of non-critical
string:
\bea
T &=& - \shalf(\pa \phi)^2 - \shalf(\pa X)^2
 + { k-1 \over \sqrt{2(k-2)}}\pa^2 \phi -
{ k-3 \over \sqrt{2(k-2)}}i \pa^2 X + T_{bc}, \nonumber\\
G^- &=& b, \nonumber\\
G^+ &=& j_{BRST}(z)-\frac{1}{k-2}\left[ \frac{k-4}{2}\pa^2 c
 +i\sqrt{2(k-2)}\pa(c\pa X)\right],
\ena
where $j_{BRST}$ is the ordinary BRST current and
$T_{bc}$ is the energy-momentum tensor of ghosts with conformal
weight $(2,-1)$:
\bea
T_{bc} &=& -2b\pa c - \pa b c \nonumber\\
&=& -\shalf\left( \sqrt{\frac{k-2}{k}}\pa u -\sqrt{\frac{2}{k}}\pa
\tau\right)^2
+\frac{3i}{2}\left( \sqrt{\frac{k-2}{k}}\pa^2 u -\sqrt{\frac{2}{k}}
 \pa^2 \tau \right).
\ena

When $k-2 = {q \over p}$, the energy-momentum tensor in (16) is that
of gravity coupled to $(p,q)$ minimal matter. We see in particular that
$k=3$ corresponds to $c=1$ string. The BRST charge obtained from
$G^+$ in (16) is the usual one, and the cohomology in the boson Fock
space is the same~\cite{LIZ,MUK,WIT2}.\footnote{If one takes the coset
representations, it may be different because of the presence of null
states~\cite{IKOS}.}
For the central charge $c<1$, however, we have to make further reduction
using Felder cohomology~\cite{FEL}.

The screening operators (4) of $SL(2,R)$ is given by
\EQ
V_- = e^{\sqrt{ 2 \over k-2} \phi}, \qquad
V_+ = e^{\sqrt{ 2(k-2)} \phi}.
\EN
We see that $V_-$ can be identified with the cosmological constant
operator. In this identification of matter and Liouville fields, we
lack the information on the origin of screening operators for the
matter field. This problem is
not present, however, for $k=3$ since screening operators are not
required for $c=1$ matter. We have thus made an explicit connection
of super $SL(2,R)_k/U(1)$ at $k=3$ and $c=1$ gravity. These results
confirm the observation of Ref.~\cite{MV,POR}.

We should note that here is an additional structure which differs
from the analysis of Ref.~\cite{MV} in this correspondence.
Namely we have another screening operator (6) which arises from our
particular bosonization (5) of $(\b, \c)$ system. The operator can
be written in the fields of $c\leq 1$ string as
\EQ
S(z) = be^{\sqrt{k-2 \over 2}i(X+i \phi)}.
\EN
When $k=3$, the zero form version of this operator is given by
\EQ
x(z) = \left( cb + {i \over \sqrt{2}}(\pa X -i \pa \phi)\right)
 e^{{i \over \sqrt{2}} (X + i \phi)},
\EN
which is exactly the state known as the Lian-Zuckerman state of ghost
number $0$~\cite{LIZ,MUK,WIT2}. Other elementary state $y(z)$ can
be obtained by $y(z) = J'_- x(z)$, with $J'_- \equiv \oint\dz
 \exp[-i\sqrt{2}X(z)]$. As is pointed out by Witten~\cite{WIT2},
these form the ground ring of $c=1$ string.

It is quite interesting to note that there is another identification
of Liouville and matter fields. When $ k-2 = - { q \over p}$, the
role of $X$ and $\phi$ interchanges: We can identify $\phi$ as $(p,q)$
matter and $X$ as Liouville field. The screening operators (4) or (18)
of $SL(2,R)$ then become those for matter field in the $(p,q)$ minimal
models:
\EQ
V_- = e^{-i\sqrt{2p/q}X}, \qquad
V_+ = e^{i\sqrt{2q/p}X}.
\EN
In this case, $k=1$ is critical in the sense that the screening
operators (21) become generators of $SU(2)$ in the matter sector.
With $X \leftrightarrow \phi$, another screening operator (19) can
be identified with BRST-nontrivial state of gravity coupled to
minimal matter and again gives one of the generators
of ground ring for $c=1$ gravity.

In this identification, the origin of the cosmological constant
operator is not clear. We have to consider the perturbation by the
cosmological constant.
The operator corresponding to the cosmological constant
\EQ
 c e^{i \sqrt{2 \over k-2} X},
\EN
is expressed by $N=2$ fields as
\EQ
e^{i \sqrt{ k \over k-2} u},
\EN
which is the chiral primary field of conformal weight
\EQ
h = { 1 \over 2} { k \over k-2} = { c \over 6}.
\EN
This state seems to have a special meaning because it is the
chiral field of the highest conformal weight, at least for
the unitary series~\cite{LVW}. (We do not know much about the
non-unitary series.) In other words, the perturbation by
the cosmological constant can be identified with that by the chiral
primary field of conformal weight $h=c / 6$. It is quite interesting to
note that this is precisely the integrable perturbation for the unitary
minimal model~\cite{EY}.

Finally let us point out that we could have defined ``$c=1$ fields"
at the outset by
\bea
{\hat\phi} &=& \frac{k-1}{2\sqrt{(k-2)}} \phi -
 \frac{k-3}{2\sqrt{(k-2)}}i X, \nonumber\\
i{\hat X} &=& \frac{k-3}{2\sqrt{(k-2)}} \phi -
 \frac{k-1}{2\sqrt{(k-2)}}i X.
\ena
The energy-momentum tensor as well as the BRST operator obtained from
(16) are then precisely those for $c=1$ matter coupled to gravity
 irrespective of the level $k$. The cosmological
constant operator (18) now takes the form
\EQ
V_- =\exp\left[\frac{k-1}{\sqrt{2}(k-2)}{\hat\phi}
 - \frac{k-3}{\sqrt{2}(k-2)}i{\hat X} \right].
\EN
The special value $k=3$ is singled out by the fact that this precisely
gives the correct operator for $c=1$ gravity.

In conclusion, we have shown that the topological algebra of twisted
super $SL(2,R)_k/U(1)$ model can be converted to that of gravity
coupled to matter field with suitable redefinition of the fields.
BRST nontrivial state appears as a screening operator under this
redefinition. It seems interesting to consider
further correspondence between these models through free field
realization.

\vs{5}
\noindent
{\it Acknowledgements}

We would like to thank Hiroshi Kunitomo for valuable discussions.
We would also like to thank T. Eguchi and S. -K. Yang for comments.

\end{document}